\documentclass[12pt,a4paper]{article}
\usepackage{times}

\usepackage[textheight=24.7cm,textwidth=18.3cm]{geometry}

\usepackage{graphicx}%
\usepackage{multirow}%
\usepackage{amsmath,amssymb,amsfonts}%
\usepackage{amsthm}%
\usepackage{mathrsfs}%
\usepackage{color}
\usepackage[backend=biber,style=nature]{biblatex}
\AtEveryBibitem{%
\ifentrytype{article}{
    \clearfield{url}%
    \clearfield{issue}%
    \clearfield{month}%
    \clearfield{URL}%
    \clearfield{eprint}
    \clearfield{issn}
}{}
}
\title{Real-time observation of picosecond-timescale optical quantum entanglement toward ultrafast quantum information processing}
\author{Akito Kawasaki,$^{1,\ast}$ Hector Brunel,$^{1,2}$ Ryuhoh Ide,$^{1}$ Takumi Suzuki,$^{1}$\\
Takahiro Kashiwazaki,$^{3}$ Asuka Inoue,$^{3}$ Takeshi Umeki,$^{3}$ Taichi Yamashima,$^{1}$\\
Atsushi Sakaguchi,$^{4}$ Kan Takase,$^{1,4}$ Mamoru Endo,$^{1,4}$ Warit Asavanant,$^{1,4,\ast\ast}$\\
and Akira Furusawa$^{1,4,\ast\ast\ast}$\\
\\
\normalsize{$^{1}$Department of Applied Physics, School of Engineering, The University of Tokyo,}\\
\normalsize{7-3-1 Hongo, Bunkyo-ku, Tokyo 113-8656, Japan}\\
\normalsize{$^{2}$Department of Physics, Ecole Normale Sup\'{e}rieure,}\\
\normalsize{24, rue Lhomond, Paris 75005, France}\\
\normalsize{$^{3}$NTT Device Technology Labs, NTT Corporation,}\\
\normalsize{3-1 Morinosato Wakamiya, Atsugi, Kanagawa 243-0198, Japan}\\
\normalsize{$^{4}$Optical Quantum Computing Research Team, RIKEN Center for Quantum Computing,}\\
\normalsize{2-1 Hirosawa, Wako, Saitama 351-0198, Japan}\\
\normalsize{To whom correspondence should be addressed; $^{\ast}$E-mail: kawasaki@alice.t.u-tokyo.ac.jp}\\
\normalsize{$^{\ast\ast}$E-mail: warit@alice.t.u-tokyo.ac.jp, $^{\ast\ast\ast}$E-mail: akiraf@ap.t.u-tokyo.ac.jp}
}
 
\date{\today}
\begin{document}
\baselineskip24pt
\maketitle

{\bf
Entanglement is a fundamental resource of various optical quantum-information-processing (QIP) applications.
Towards high-speed QIP system, entanglement should be encoded in short wavepackets.
We report real-time observation of ultrafast optical Einstein-Podolsky-Rosen (EPR) correlation at a picosecond timescale in a continuous-wave (CW) system.
Optical phase-sensitive amplification using 6-THz-bandwidth waveguide-optical-parametric amplifier enhances the effective efficiency of 70-GHz-bandwidth homodyne detectors, mainly used in 5th-generation telecommunication, enabling its use in real-time quantum-state measurement.
While power measurement using frequency scanning, i.e., optical spectrum analyzer, is not performed in real-time, our observation is demonstrated through real-time amplitude measurement and can be directly employed in QIP applications.
Observed EPR states show quantum correlation of 4.5 dB below shotnoise level encoded in wavepackets with 40-ps period, equivalent to 25-GHz repetition---$\bf{10^3}$ times faster than previous entanglement observation in CW system.
The quantum correlation of 4.5 dB is already sufficient for several QIP applications, and our system can be readily extended to large-scale entanglement.
Moreover, our scheme has high compatibility with optical communication technology such as wavelength-division multiplexing, and femtosecond-timescale observation is also feasible.
Our demonstration is paradigm shift in accelerating accessible quantum correlation, the foundational resource of all quantum applications, from the nanosecond to picosecond timescale, enabling ultra-fast optical QIP.

}
% Quantum entanglement is fundamental resource for quantum information processing (QIP) applications. 
% Continuous-variable (CV) optical system is a promising QIP platform, enabling the deterministic generation of large-scale entangled states and ultrafast clock speeds, in principle, up to terahertz order.
% Currently, however, the speed is limited by the homodyne detectors (HDs) up to tens of megahertz.
% The commercially available HDs with 100-GHz bandwidth are mostly limited to classical technology due to high electric noise and low quantum efficiency, making it incompatible to Quantum measurement.
% Previous research suggests that adding phase-sensitive amplification (PSA) could make them compatible to quantum measurement. This technique, however, has been limited to the single-mode cases due to the challenges arising from synchronized phase-control unique to the multimode states.
% We develop a phase-locking method for broadband homodyne measurement with PSA and apply this technique on a two-mode entangled state. 
% We observe Einstein-Podolsky-Rosen correlation up to 4.7 dB below shotnoise with ~60 GHz bandwidth---$\bf{10^3}$ times broader than conventional method. The correlation level is also sufficient for some QIP applications.
% Our phase-locking technique can be applied to larger entangled-state generation and expands the boundary of broadband entanglement generation and measurement, paving the way toward ultrafast CV-QIP with gigahertz to terahertz clock frequency.}

\section{Introduction}
Quantum entanglement is a fundamental concept of quantum mechanics \cite{PhysRev.47.777,schrdinger1935}, and also a basic resource for various applications of quantum information processing (QIP), such as quantum computation \cite{doi:10.1137/S0097539795293172,doi:10.1098/rspa.1985.0070}, quantum communication \cite{PhysRevLett.69.2881,PhysRevLett.81.5932,Gisin2007}, and quantum cryptography \cite{bennett2014quantum,PhysRevLett.67.661,Pirandola:20}. 
Since light has a carrier frequency of hundreds terahertz and is compatible with the highly developed optical communication technology, it is attracting attention as a promising platform for realizing these QIP applications with high clock frequencies.
In particular, generation of large-scale quantum entangled states called cluster states have already been reported in continuous-variable (CV) optical system \cite{doi:10.1126/science.aay2645,doi:10.1126/science.aay4354}. Cluster states are resources for large-scale measurement-based quantum computation (MBQC) \cite{RevModPhys.77.513,takeda2019toward} and quantum communication \cite{RevModPhys.77.513}, and such large-scale CV quantum entanglement is expected to be a basic technology towards optical QIP.
One of the main directions in the development of CV optical quantum technology is to speed up the system clock frequency. This strategy is also taken in the classical optical communication to communicate and process more information in a single optical channel. As quantum states are defined in a temporal wavepacket in CV optical QIP, the time width of the wavepacket has to be reduced to accelerate the system clock. Note that, the time width of wavepacket is determined by the system lock only in continuous wave (CW) systems, whereas the repetition rate does in the pulsed systems.

\begin{figure}[ht]
\centering
 \includegraphics[width=18cm]{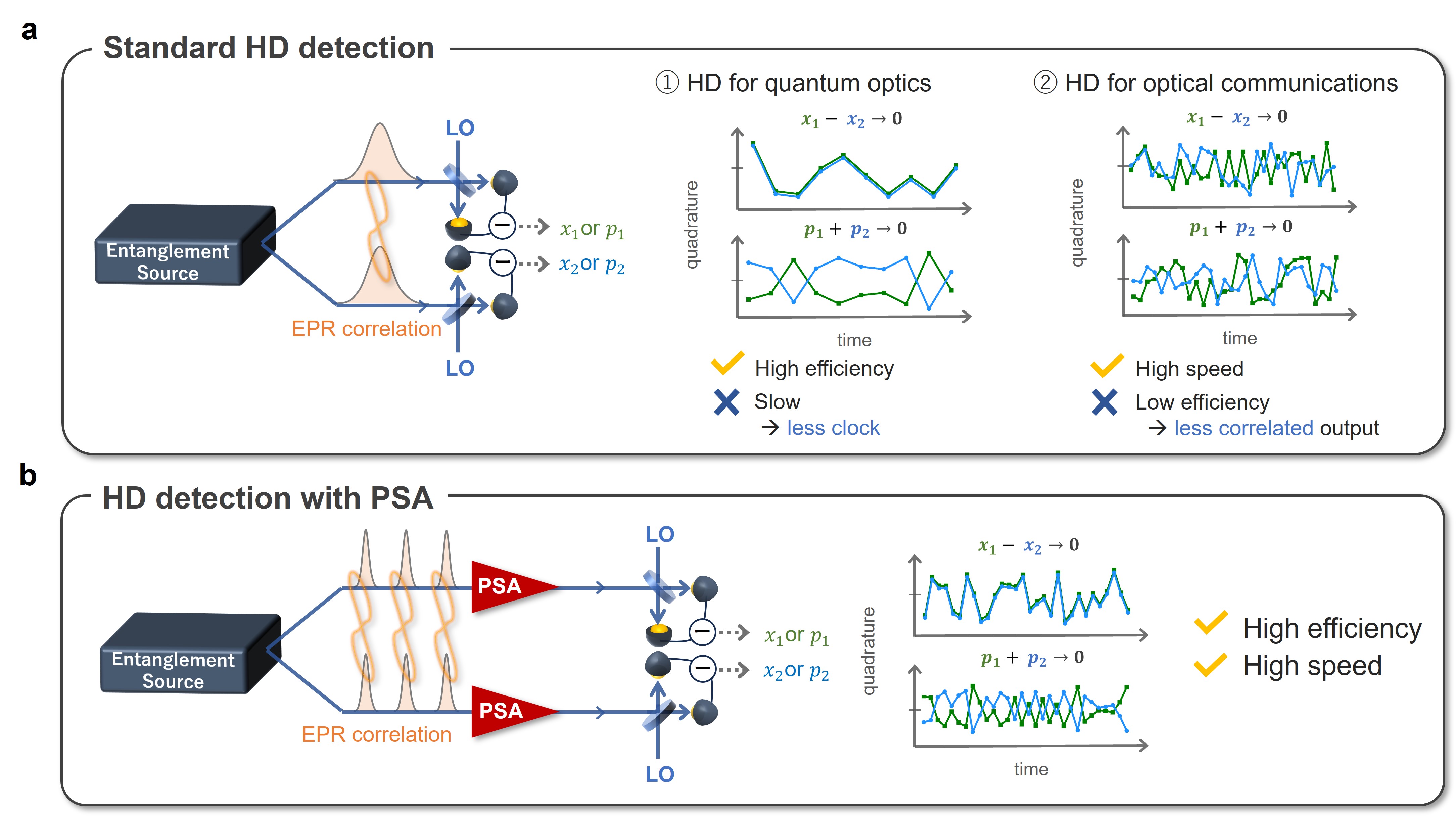}
 \caption{\textbf{Schematic diagram of measurement of EPR correlation using homodyne measurement.} (a) Without phase-sensitive amplifiers. (b) With phase-sensitive amplifiers. Standard homodyne measurement schemes used in quantum optics or optical communications cannot simultaneously attain high-efficiency and high-speed measurement, making them unsuitable for measurement of broadband quantum entangled states. By adding a phase-sensitive amplifier as a preamplifier with 0-dB noise figure, low-efficiency and high-speed homodyne can be used. HD, homodyne detection; PSA, phase sensitive amplification; LO, local oscillator.}

\label{gaiyoh}
\end{figure}

An example of the approach to shorten the wavepacket is the single-path waveguide optical parametric amplifier (OPA) \cite{doi:10.1126/science.abo6213,Chen:22,10.1063/5.0144385}. This OPA has THz-bandwidth, and in principle, it can generate squeezed light defined in sub-picosecond timescale wavepacket, attracting attention as a next-generation ultrafast light source. 
On the other hand, the usages of quantum entanglement, such as CV-MBQC or quantum communication, are based on "measurement" of wavepacket and "feedforward" based on the measurement results. 
Therefore, in addition to "state pareparation" with short wavepackets, speeding up the "measurement" is also critical for CV optical QIP. 
In particular, as quantum information in CV-QIP is encoded in the phase space, the measurement of our interest is the real-time amplitude measurement.
This measurement is the homodyne measurement \cite{Yuen:83}, which has been a basic optical technique for measuring the real-time amplitude, i.e., the quadrature operators.
Note that power measurement using frequency scanning, such as optical spectrum analyzers with terahertz bandwidth, is not performed in real time and irrelevant to actual optical QIP applications.
Figure \ref{gaiyoh} (a) shows the trade-off between the speed and the efficiency of the conventional homodyne measurement. To ensure that we extract correct quadrature values, the efficiency of the measurement must be high and this limits the other specification of the photodiode. Experimentally, the high-efficiency (more than 95\%) homodyne detector has been realized with the frequency bandwidth of 200 MHz at best \cite{Kawasaki:22}. On the other hand, with recent developments of optical communication technologies for 5-th or 6-th generation communication systems, balanced detectors with bandwidth of up to $113$ GHz are now available \cite{7968460}. These high-speed homodyne detectors, however, are developed for coherent lights that are robust to loss and the high-frequency bandwidth is achieved by trading the efficiency ($\sim50\%$) for bandwidth, making them incompatible with quantum measurement. This trade-off relationship has limited the system clock of the quantum state generation and operation \cite{Kawasaki:22,Darras2023} to tens of nanosecond timescale at least which is much slower than the classical optical communication systems whose clock cycle is sub-nanosecond. Especially, to the best of our knowledge, the shortest record for entanglement observation involves 40-ns wavepackets \cite{doi:10.1126/science.aay2645}.

To overcome this trade-off limitation, phase-sensitive amplifier (PSA) \cite{PhysRevD.26.1817} has attracted attention recently as a method for applying this low-efficiency and high-speed homodyne detector to quantum-optics technologies. Figure \ref{gaiyoh}(b) shows the schematic diagram of this approach. Here, the PSA acts on the quantum states prior to the homodyne measurement. The role of the PSA is to act as a preamplifier with 0-dB noise figure in principle for quadrature measurements, enabling high-precision quantum measurements, even when the efficiency of the subsequent homodyne detectors is low. PSA in optical system can be achieved by parametric amplification and has been demonstrated experimentally \cite{Shaked2018,doi:10.1126/science.abo6213,Takanashi:20}. Most of these demonstrations, however, measure the quadrature "power" and not their real-time amplitude, making them less relevant to actual QIP applications. A demonstration of measurement of real-time quadrature amplitude in this manner was done using a low-loss high-gain OPA, and the real-time quadrature values of the squeezed light have been measured up to 43 GHz bandwidth \cite{10.1063/5.0137641}. In all of these previous demonstrations, regardless of whether it is "power" or "amplitude" measurement, the preparation and measurement are limited to squeezed light defined in the single mode and never extended to quantum entanglement, i.e., the multimode quantum state, which are the fundamental resources for all QIP applications.

In this research, we demonstrate generation and real-time measurement of ultrafast Einstein–Podolsky–Rosen (EPR) state, the most primitive CV quantum entangled state by using homodyne measurement with PSA.
We observe a quantum correlation of more than 4.5 dB below the shotnoise level encoded in a wavepacket with 40-ps period, equivalent to 25-GHz repetition---$10^3$ times faster than conventional quantum entanglement observation. 
The observed EPR states have auto-correlation with 20-ps time width, meaning that we can introduce tens-of-picosecond wavepacket to exploit their quantum correlation.
To the best of our knowledge, demonstration of EPR state generation using a waveguide OPA has never been done before and the correlation level observed here is already sufficient for some applications in optical CV-QIP \cite{PhysRevA.60.2752,doi:10.1126/science.aay2645}. 
For this experiment, we develop a new phase-locking method for homodyne measurements with PSA on two-mode states. Our method is not limited to two-mode states, but can be readily extended to the multimode quantum states and large-scale quantum entangled state previously demonstrated \cite{doi:10.1126/science.aay2645,doi:10.1126/science.aay4354}. Moreover, our scheme has a high compatibility with optical communication technology; femtosecond-timescale observation is also feasible by using wavelength division multiplexing. Our demonstration is a milestone toward ultrafast QIP applications via quantum entanglement.

\section{Result}
\subsection{Homodyne detection with phase sensitive amplification}

\begin{figure}[ht]
\centering
 \includegraphics[width=18cm]{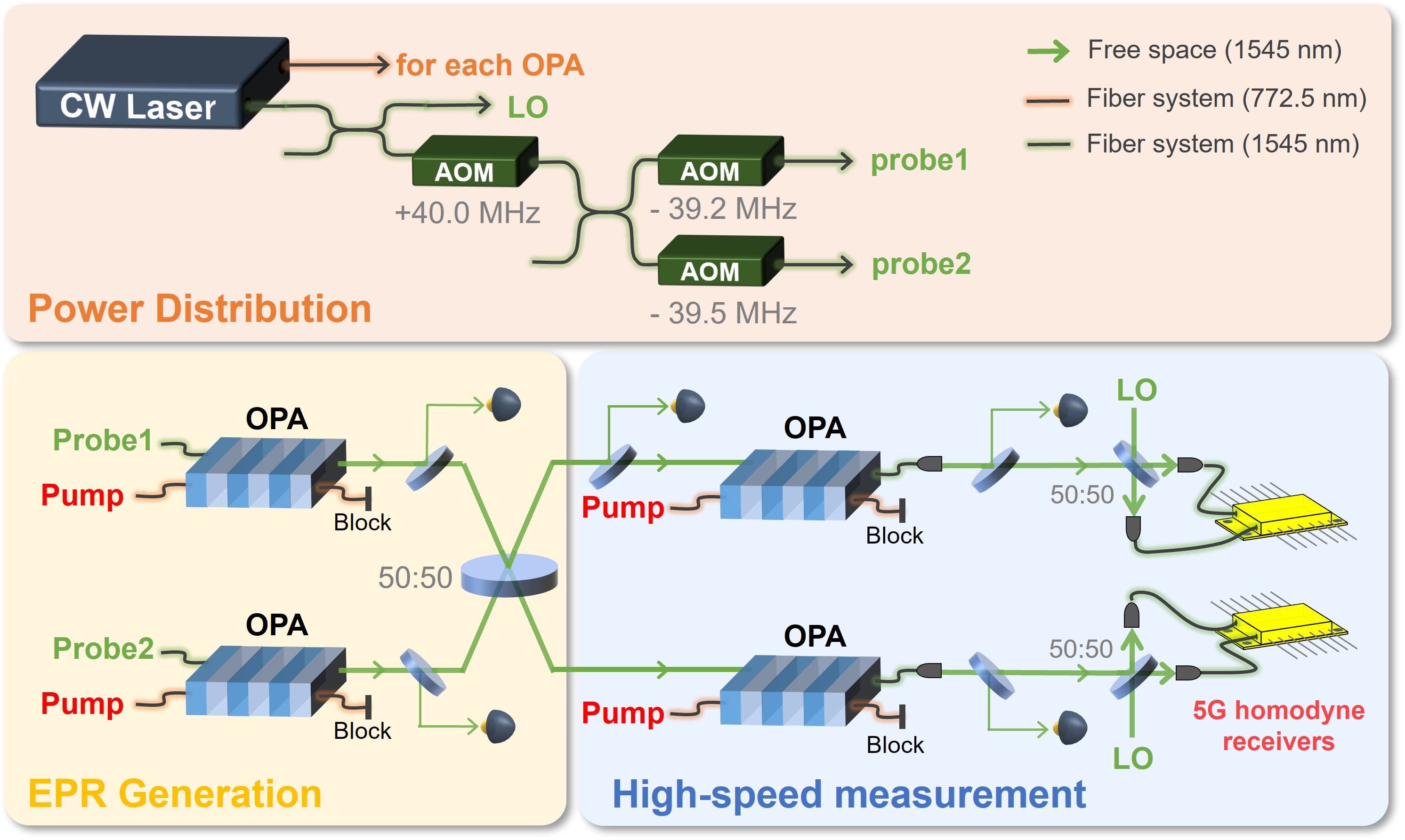}
 \caption{\textbf{Experimental setup.} The fundamental laser is a continuous-wave laser operating at the wavelength of $1545$ nm with the second harmonic at the wavelength $772.5$\,nm also generated. EPR states are generated by interfering the squeezed lights from the two generation OPAs on a half beamsplitter. Each mode of the EPR states is amplified by the measurement OPAs and is sent to the homodye detectors. AOM, acousto-optic modulator; LO, local oscillator.}
 \label{setup}
\end{figure}
Figure \ref{setup} shows the schematic of the experimental setup and the details are described in Method section.
The squeezed-light sources and the PSAs in this experiment are periodically-poled-lithium-niobate (PPLN) waveguide OPAs \cite{doi:10.1063/1.5142437} developed by our group. 
First, squeezed lights are generated in the two generation OPAs and they are interfered in such a the relative phase that the directions of squeezing are orthogonal. This results in a two-mode squeezed state which approaches the EPR state in the infinite squeezing limit.
Formally, let us consider the correlation of the quadrature amplitudes, which will be denoted by $\hat{x}$ and $\hat{p}$ satisfying $[\hat{x},\hat{p}]=i$. If we assume that the two squeezed lights have the same squeezing parameter $r$, then the quadrature amplitudes of the output modes after the interference satisfy $\hat{x}_1-\hat{x}_2 = \sqrt{2} e^{-r}\hat{x}_{\rm vac},\,\hat{p}_1+\hat{p}_2 = \sqrt{2} e^{-r}\hat{p}_{\rm vac}$, where the indices 1,2 are the mode indices and $\hat{x}_{\rm vac}$, $\hat{p}_{\rm vac}$ represent quadrature amplitudes of the vacuum states. At the infinite squeezing limit, i.e., $r$ approaching infinity, the quadrature amplitudes $\hat{x}$ ($\hat{p}$) will have perfect (anti-)correlation.

After the generation of the two-mode squeezed states, each mode enters the PSA at the measurement OPAs. The role of the measurement OPAs is to amplify the quadrature. This amplification acts only on a single phase of the quadrature, allowing amplification with 0-dB noise figure in principle, which reduces the degradation of the EPR correlation due to the losses of the homodyne detectors. On the contrary, the quadrature amplitude that is orthogonal to the amplified phase will be compressed, making it more vulnerable to noise. This does not affect the final measurement if we are measuring quadrature amplitudes at the amplified phase. Note, however, that this means that the relative phase between the amplified quadrature and the local oscillator (LO) light of the homodyne detector will be important as they determines the efficiency of the pre-amplification with PSA in measuring the EPR correlation. The efficiency of the measurement system which consists of the measurement OPAs and the high-speed homodyne detector can be expressed as \cite{10.1063/5.0137641}
\begin{equation}
\label{PSAeq}
\eta_{\rm meas} = \frac{\eta_{\rm OPA} \eta_{\rm HD}}{\eta_{\rm HD} +\frac{1-\eta_{\rm HD} }{G}},
\end{equation}
and the total efficiency of experimental setup is written by
\begin{equation}
\label{totaleq}
\eta_{\rm total} = \eta_{\rm state} \eta_{\rm meas}. 
\end{equation}

Here, $\eta_{\rm OPA}$, $\eta_{\rm HD}$, $\eta_{\rm state}$ and $G$ are the efficiency of the OPA used for PSA, efficiency of homodyne detection, efficiency of state preparation, and the amplification gain of the PSA, respectively. 
From Eq.\eqref{PSAeq}, we see that when there is no amplification $(G=0\,\rm{dB})$, the measurement efficiency is $\eta_\textrm{OPA}\eta_\textrm{HD}$, while in the limit of large amplification gain, the measurement efficiency becomes $\eta_\textrm{OPA}$. Thus, the amplification suppresses the effect of the optical losses of the homodyne detector.

%Equation \ref{PSAeq} shows the effective loss after amplification is suppressed by amplification gain $G$.

\subsection{Raw data from real-time 110GHz oscilloscope}

\begin{figure}[ht]
\centering
 \includegraphics[width=16cm]{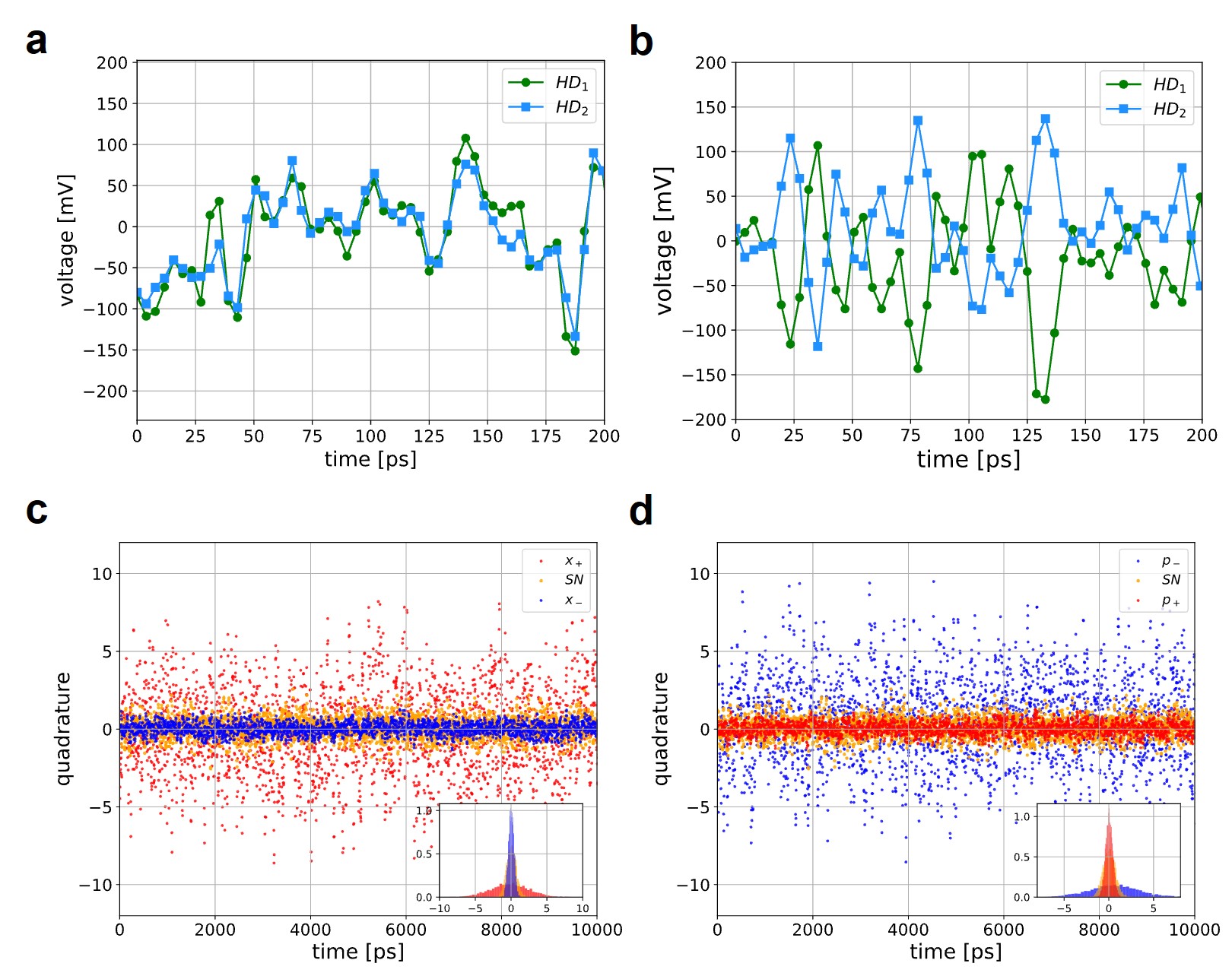}
 \caption{\textbf{Real-time quantum correlation signal.} (a,b) Real-time output from the two homodyne detectors for $x$ and $p$ quadrature measurement, respectively. We can see the correlation and anti-correlation of the quadrature values between two modes in the picosecond time scale. (c,d) The comparison of shot noise level and noise power of the linear combinations of quadrature amplitudes given by $x_+=(x_1+x_2)/\sqrt2,\,x_-=(x_1-x_2)/\sqrt2,\,p_+=(p_1+p_2)/\sqrt2$, and $p_-=(p_1-p_2)/\sqrt2$ in time domain. Both results show noise power suppression below shot noise level for one of the quadrature amplitudes, indicating our successful observation of quantum correlation.}
 \label{time}
\end{figure}
Figure \ref{time} shows the real-time signal measured by the oscilloscope.
The real-time signals clearly show correlation and anti-correlation in $x$ and $p$ quadrature amplitudes on the timescale of picosecond. It is notable that this measurement is real-time amplitude measurement and these outputs can be directly utilized in QIP operations. The timescale of the real-time signal is limited by the sampling rate (256 GSa/s) and the bandwidth (113 GHz) of the oscilloscope (see Method). In the ideal case, the quadrature would be perfectly correlated or anti-correlated and the imperfections observed here are due to factors such as finite squeezing and optical losses in the system. Quantitative evaluation and verification of quantum entanglement between these two modes can be done by calculating the noise power of the linear combinations of quadrature amplitudes given by $x_+=(x_1+x_2)/\sqrt2,\,x_-=(x_1-x_2)/\sqrt2,\,p_+=(p_1+p_2)/\sqrt2$ and $p_-=(p_1-p_2)/\sqrt2$, and comparing them to the shotnoise level.

\subsection{Analysis on quantum correlation in time domain}
\begin{figure}[ht]
\centering
 \includegraphics[width=18cm]{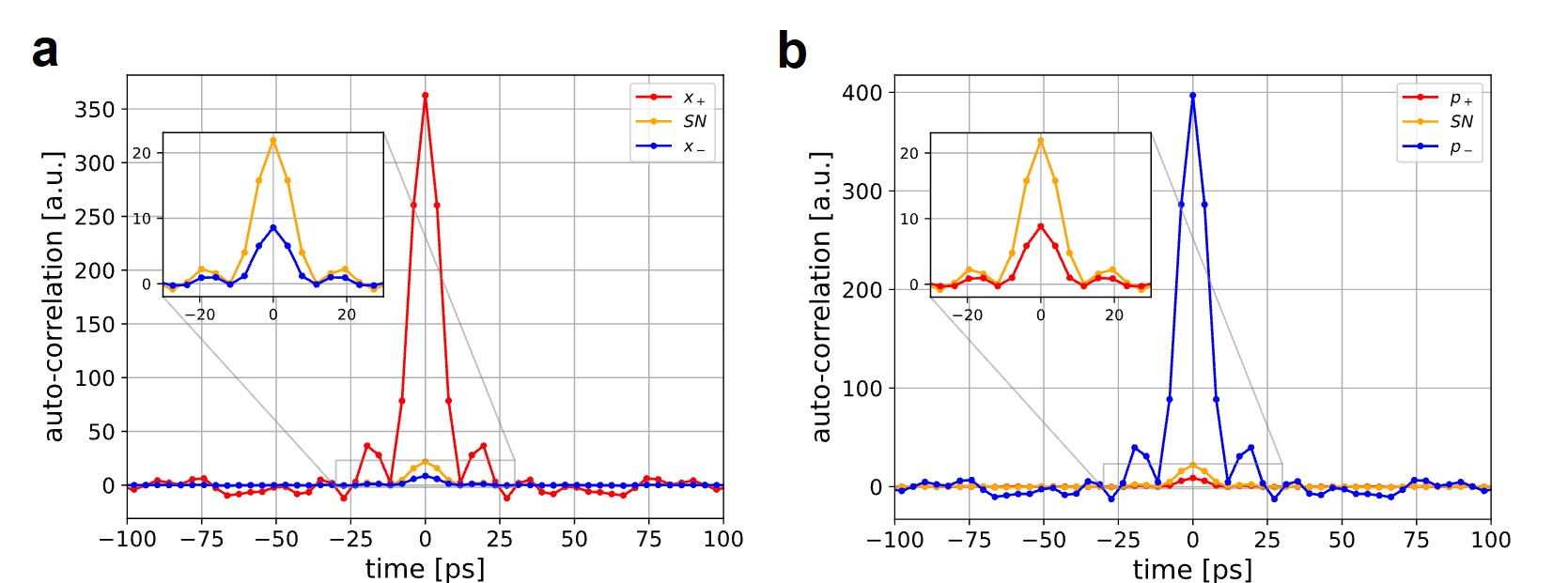}
 \caption{\textbf{Analysis on quantum correlation in time domain.} (a,b) Auto-correlation functions of $x$ and $p$ quadrature amplitudes, respectively. We can find $x_-$ and $p_+$ are suppressed below shotnoise level, indicating successful observation of qunatum correlation. The auto-correlation functions have time width of about 20 ps, which represent the lower limit of the wavepacket width where entanglement can be encoded.}
 \label{autocorr}
\end{figure}
Figure \ref{autocorr} (a,b) show the auto-correlation functions of $x_\pm$ and $p_\pm$, respectively.
Auto-correlation function $I_{\rm cor}^{\pm}(\tau)$ is defined as $I_{\rm cor}^{\pm}(\tau) = \langle x_\pm(t)x_\pm(t+\tau) \rangle_t$ and characterizes quantum correlation in time domain. 
Here, $\langle...\rangle_t$ represents the temporal average over time $t$, indicating the expectation value with respect to $t$. 
From Fig. \ref{autocorr} (a,b), we can see that both $x_-$ and $p_+$ are below the shotnoise level around $\tau = 0$.
The values at $\tau = 0$ represent the noise power of the quadrature amplitudes shown in Fig. \ref{time} (c,d), and $x_-$ and $p_+$ are both 4.0 dB below the shotnoise level at $\tau = 0$, resepectively. 
These levels of quantum correlations satisfy the inseparability criterion for CV entangled states \cite{PhysRevLett.84.2722,PhysRevLett.84.2726,PhysRevA.67.052315}, meaning that we have succeeded in the generation and measurement of quantum entanglement.
Moreover, Fig. \ref{autocorr} (a,b) show that the time widths of auto-correlations are around 20 ps.
When considering the use of entanglement in QIP applications, it is necessary for adjacent wavepackets to be independent and uncorrelated. 
Therefore, the width of the auto-correlation function sets a lower limit on the clock cycle of QIP system.
Both of broadband squeezed light source and high speed measurement enable this ultrafast quantum correlation observation with tens-of-picosecond timescale.
As an example, we calculate quantum correlation defined in a wavepacket. (See supplymentary \cite{supp} for details of quantum-correlation analysis using wavepacket).
The shape of the wavepacket is known to be useful for quantum computaion \cite{10.1063/1.4962732} and the same shape as one employed in conventional quantum entanglement observations \cite{doi:10.1126/science.aay2645}. 
Here, only the time width is shortened by a factor of $10^3$ from the conventional research of 40 ns \cite{doi:10.1126/science.aay2645}, which is the previous record of the shortest time-scale observation of entanglement.
The obtained quantum correlations of $x_-$ and $p_+$ are below the shot noise level by 4.7 dB and 4.5 dB respectively, indicating successful observation of quantum correlation with $10^3$ times high-speed system than conventional research.

\section{Discussion}
In this experiment, we experimentally demonstrate the real-time observation of high-speed EPR states with qunatum correlations of more than 4.5 dB.
%In this experiment, real-time observation of EPR states over a 60 GHz bandwidth with correlations of up to 4.7 dB was demonstrated.
The time width of wavepacket we use is more than $10^3$ times faster than the previous quantum entangled state obsevation \cite{doi:10.1126/science.aay2645}.
The correlation level of 4.5 dB can already be used in some applications; it exceeds the 3 dB required for entanglement swapping with unit gain \cite{PhysRevA.60.2752}, and 3 dB \cite{doi:10.1126/science.aay4354} or 4.5 dB \cite{doi:10.1126/science.aay2645} required for the verification of 2D cluster states toward CV-MBQC.

Further development of this research has three directions: "larger scale", "higher speed" and "stronger correlation".
Large-scale entanglement is an important aspect in the context of MBQC and quantum communication.
The synchronization scheme for phase-sensitive amplified homodyne measurements developed in this research can be directly applicable to large-scale cluster state generation \cite{doi:10.1126/science.aay2645,PhysRevA.97.032302} using only linear optical elements.
Therefore, this research is the first step paving the way toward high speed generation of large-scale cluster states.

Regarding higher speed, electrical system is the current limiting factor. In particular, the 1.85-mm connectors (V-connectors) of the electrical components in the measurement system limit the bandwidth of the measurement up to 60 GHz and the time scale of EPR correlations to tens of picosecond. The generation part, on the other hand, is much broader as the bandwidth of the squeezed light source is about 6 THz \cite{doi:10.1063/1.5142437}. Therefore, even with the same setup, higher speed EPR correlations can be measured by increasing the bandwidth of the measurement system. Even with current commercially-available optical communication technology, we expect a bandwidth extension of up to about 100 GHz to be readily achieved. The full usage of the THz-bandwidth quantum entanglement and femtosecond-timescale measurement might also be achieved in the future by combining the wavelength-division multiplexing technique \cite{1634536}.

Finally, we discuss possibility toward EPR state generation and measurement with stronger correlation. 
First, the efficiency of EPR state preparation $\eta_\textrm{state}$ and measurement $\eta_\textrm{meas}$ can be estimated from the efficiency of each individual component as $\eta_\textrm{state} = 94\%$, $\eta_\textrm{meas}(G=0\,\textrm{dB}) = 19\%$, and $\eta_\textrm{meas}(G=25\,\textrm{dB})=76\%$. These values match the result of the experimental fitting (see Method for details).
Towards further purification of the EPR states, the coupling from free space to the waveguide OPAs, which is the largest imperfection in the measurement system $\eta_{\textrm{meas}}$, is expected to be significantly improved. By optimizing the alignment, the measurement efficiency $\eta_\textrm{meas}$ can be increased from the current efficiency of $80\%$ to more than $95\%$ \cite{10.1063/5.0137641}. Furthermore, as the intrinsic loss of the OPA, for both state preparation and measurement, is roughly proportional to the waveguide length, shortening the crystal length will improve the efficiency at the cost of requiring higher pump power.

In conclusion, this research demonstrates real-time observation of picosecond-timescale two-mode quantum entanglement. The observed EPR
states show the quantum correlation of more than 4.5 dB with 
a wavepacket of 40-ps period. Our scheme will be the fundamental technology paving the way to a wide range of ultrafast optical QIP applications.

\section{Method}
Figure \ref{setup} shows the experimental setup.
The fundamental laser is a continuous wave laser at the wavelength of $1545$ nm and $772.5$ nm. 
We utilize PPLN waveguide OPAs \cite{doi:10.1063/1.5142437} we developed as both a squeezed light source and a phase sensitive optical amplifier. 

The details of the $\eta_\textrm{state}$ and $\eta_\textrm{meas}$ are as follows. The efficiency of the state preparation $\eta_\textrm{state}$ is determined by the propagation efficiency ($98\%$), the escape efficiency of the generation OPAs ($96\%$) \cite{doi:10.1063/5.0063118}, and the interferometric visibility at the first beam splitter ($99.5$\%). On the other hand, the efficiency of the measurement system $\eta_\textrm{meas}$ without amplification ($G=0\,\textrm{dB}$) is determined by the coupling efficiency from free space to fiber at the measurement OPAs ($90\%$) and the homodyne detector ($90\%$), interferometric visibility at the homodyne measurement ($99.6\%$), and efficiency of the broadband homodyne detector ($36\%$). We numerically fit the efficiency of the system using two parameters: the efficiency before and after light enter the measurement OPAs. From the experimental parameters, we expect this to be around $70\%$ and $20\%$, respectively. The fitting results are $68\%$ and $16\%$ which are in agreement with the experimental parameters (See Supplementary material \cite{supp} for the details of the fitting). The pump power of the generation and the measurement OPAs are 200 mW and 800 mW (measured after OPAs), respectively. At this parameter, the gain of the measurement OPAs is about 25 dB. The LO power is set to 45 mW, resulting in the clearance between the shotnoise and circuit noise to be more than 15 dB up to 20 GHz, and more than 10 dB up to 60 GHz.

The details of measurement system are as follows.
The output of homodyne detector (BPDV3120R, Finisar) is amplyfied with 22 dB gain by the amplifier (M804C, SHF). After amplification, the signal is devided into two parts, one for oscilloscope (Infiniium UXR-Series, Keysight) and the other for phase locking system.
The bandwidth of the homodyne detectors is $70$ GHz and the efficiency is 36\%. 
The bandwidth of amplifiers is from $90\,$kHz to $66\,$GHz.
The bandwidth of the oscilloscope is $113$ GHz and the sampling frequency is $256$ GSa/s.
We take 5000 frames of data with 5121 points each.
Each components are connected by $1.85$-mm connector (v-connector), and their bandwidth of $66$ GHz limits the bandwidth of whole measurement.

In this experiment, we implemented 7 phase lockings: the phases of the two squeezed light, the interference phase between the two squeezed light, the amplification phases at the measurement OPAs, and the phases of the LO light at the homodyne detectors. 
We use strong classical probe light injected from the generation OPAs in order to implement these phase locking.
Phase locking is performed by detuning the two probe light from the fundamental frequency by $0.8$, and $0.5$ MHz, and observing the beat signal.
Probe light is strong classical light and it can disturb quantum measurement, therefore we separate phase control and measurment timing in time domain.  
The frequency detuning and the on-off switching of probe light is conducted by three AOMs.
In this experiment, 1 cycle is $400\,{\rm \mu s}$, of which the control phase is $360\,{\rm \mu s}$ and the measurement phase is $40\,{\rm \mu s}$.

\section*{Acknowledgement}
The authors are grateful to R. Nehra for useful discussions. This work was partly supported by Japan Science and Technology (JST) Agency (Moonshot R\&D) Grant No. JPMJMS2064, the UTokyo Foundation, and donations from Nichia Corporation. W.A. acknowledges the funding from Japan Society for the Promotion of Science (JSPS) KAKENHI (No. 23K13040). M. E. acknowledge the funding from JST (JPMJPR2254). A.K., R.I., and T.S. acknowledge supports from Forefront Physics and Mathematics Program to Drive Transformation (FoPM), a World-leading Innovative Graduate Study (WINGS) Program, the University of Tokyo. A.K. acknowledges financial support from Leadership Development Program for Ph.D (LDPP), the University of Tokyo. M.E. and W.A. acknowledge supports from the Research Foundation for Opto-Science and Technology.

\section*{Author contributions}
W.A. conceived the project. A.K. led the experiment with supervision from W.A., A.S., K.T., M.E., and A.F.. A.K., H.B., R.I., T.S., and W.A. built the optical setup. A.K., W.A., H.B., R.I., T.S., T.Y., A.S., K.T, and M.E. discuss and design the experimental system and control system. W.A. prepared the high-speed electrical components and homodyne detectors, and A.K. and H.B. setup the electrical systems. A.K. and R.I. wrote the programming code for collecting the experimental data. A.K. analyzed the data with W.A. doing a separate analysis for cross checking. T.K., A.I., and T.U. provided the OPAs used in the experiment. A.K. wrote the manuscript with assistance from W.A. and all the coauthors.

\section*{Competing interests}
Authors declare no competing interests.

\section*{Data and materials availability}
All data are available either in the manuscript or in the supplementary material.

%\bibliographystyle{junsrt}
%\bibliography{ref}
\printbibliography

\end{document}